# Fiber-based optical trapping of Yeast Cells as Near-field Magnifying Lenses for Parallel Subwavelength imaging

CHUNLEI JIANG,[1,][*] HANGYU YUE,[1] BING YAN,[2,3], TAIJI DONG[1], XIANGYU CUI[4], ZENGBO WANG[2,][*]

[1]*College of Electrical and Information Engineering, Northeast Petroleum University, Daqing 163318, China*
[2]*School of Computer Science and Electronic Engineering, Bangor University, Dean Street, Bangor, Gwynedd, LL57 1UT, UK*
[3]*Center of Optics Health, Suzhou Institute of Biomedical Engineering and Technology, Chinese Academy of Sciences, No. 88 Keling Street, Suzhou Jiangsu, 215163, China*
[4]*College of Computer and Information Technology, Northeast Petroleum University, Daqing 163318, China*
*\*Corresponding author: z.wang@bangor.ac.uk, jiangchunlei_nepu@163.com*

**Subwavelength imaging by microsphere lenses is a promising label-free super-resolution imaging technique. There is a growing interest to use live cells to replace the widely used non-biological microsphere lenses. In this work, we demonstrate the use of yeast cells for such imaging purpose. Using fiber-based optical trapping technique, we successfully trapped a chain of yeast cells and bring them to the vicinity of imaging objects. These yeast cells work as near-field magnifying lenses and simultaneously pick up the sub-diffraction information of the nanoscale objects under each cell and project them into the far-field. Blu-ray disc of 100 nm feature can be clearly resolved in a parallel manner by each cell, thus effectively increasing the imaging field of view and imaging efficiency. Our work will contribute to the further development of more advanced bio-superlens imaging system.**

The ability to resolve micro and nano scale objects in high-resolution is crucial to the development of modern science and technology, especially for material science and biomedical research. Several imaging techniques are widely used, including electron and photon-based microscopy techniques such as scanning electron microscopy (SEM) and optical microscopy. The SEM has sub-nanometer ultra-high resolution, but may cause damage to the sample [1]. Optical microscopy is ideal for non-invasive imaging, but the resolution of a conventional microscope is limited by Abbe diffraction law to about half the incident wavelength, at about 200-250 nm for a white light microscope [2]. Near-field Scanning Microscope (NSOM) appeared as the first super-resolution optical imaging system, which scans the sample surface by a near-field tip collecting the evanescent waves. Its resolution can reach 20nm but its scanning efficiency is low [3-5]. Super-resolution optical microscopy has been an active research field in the past decades. Many super-resolution techniques have been proposed and demonstrated, including photon activated localization microscopy(PALM) [6], stimulated emission depletion microscopy (STED) [7], structured illumination microscopy (SIM) [8], and more [9]. Although these technologies have greatly improved optical imaging resolution, they are commercially expensive and involve complex sample labeling or image processing algorithms and may only work under narrowband laser excitation. It is therefore highly desirable to develop low-cost, label-free, real-time and easy-to-implement optical super-resolution imaging solutions which can work under conventional white lighting. The advent of microsphere super-resolution technique has emerged as a simple yet effective solution to such need.

The microsphere lens imaging technique is a method for subwavelength imaging by introducing a microscale spherical particle as add-on magnifying lens in an ordinary optical microscope. The microsphere lens was placed in the close vicinity of the imaging objects, with nanoscale information encoded in evanescent waves being scattered by the microsphere and converted into propagating waves that reached the far-field. In 2011, Wang *et al.* used this technique for the first time to resolve 50 nm features in plasmonic samples (hexagonal array with 50 nm holes separated 50 nm apart) under white light illumination [10], with calibrated resolution of $\lambda/6 - \lambda/8$ [9]. The field has since been grown rapidly. For example, super-resolution imaging of adenovirus and nanoscale samples by submerging microspheres in solution was soon developed [11-15]. To further improve the imaging quality of the microspheres, the effects of various parameters on the imaging performance, including diameter [16], immersion fluid [17], immersion mode [18-22] were widely studied. In these studies, the microspheres are randomly distributed on the sample surface at different locations, resulting in the inability to achieve complete imaging of the sample, this has greatly limited the use of this technique. L.Liu *et al.* proposed a non-invasive rapid scanning microscope by attaching a microsphere lens to an AFM tip and scan over the sample surface to obtain a stitched whole image. They applied the developed scanning nanoscope to observe various nanochip samples as well as submembrane structures in cells with great details in super-resolution over a large surface area [23]. Other scanning mechanism including attaching microsphere to a glass pipette [24], using swimming microrobot [25] and encapsulating microsphere inside solid film [26]. In attempts to finding naturally-available optical superlens, Wang and others searched for biological superlens for the first time from the natural world in 2016, they used spider silk as near-field superlens for subwavelength imaging [27]. B. Li *et al.* used the optical tweezer technique [28, 29] to trap and manipulate single cell as bio-microsphere lens for near-field nano imaging, such development opens the door to develop all-bio super-resolution imaging system by using just cells [30]. However, since the technique only used one single cell, its imaging field of view and imaging efficiency are low; it is highly desirable to develop a more efficient cell-based super-resolution imaging system. To address this issue, a new optical imaging system based on optically trapped chain of yeast cells is constructed in this paper. The imaging system uses fiber optic tweezers technique to trap and manipulate yeast cells in solution, and then use them as a chained biological microsphere superlens to image underlying nano structures, parallelly with each cell working as a superlens. The new system has boosted the imaging speed and efficiency by at least three times, with potential to scale it by using more trapping beams.

A schematic diagram of the experimental setup is shown in Fig. 1(a). All experiments were carried out in an aqueous background solution (water, $n_b$=1.33) under an optical microscope, equipped with an CCD camera, an objective lens (40×, NA=0.65), and a broadband LED illumination source (central wavelength: $\lambda$=550 nm). Such microscope system has diffraction-limited imaging resolution of approximately 315 nm (calculated by: $\lambda/2n_b.NA$) when considering the water immersion limit, which makes it able to direct image DVD disc sample (340 nm groove separated 400 nm apart) while insufficient to image Blu-ray Disc (BD) sample (100 nm groove separated 200 nm apart). A laser with 980 nm wavelength was used as fiber trapping source. The illumination source is diagonally adjusted by a collimating lens and an optical diaphragm to improve the overall image quality. As shown in Fig. 1(b), a commercial optical fiber is drawn by heating to make a tapered optical fiber for optically trapping of the yeast cells as imaging lenses. The fiber probe is fixed on the manipulator and the position of the trapped yeast cells can be controlled by moving the tapered fiber with the manipulator. To reduce undesired imaging distortion, we deliberately choose cells with smooth surface and near-spherical shape as imaging lenses, such as those shown in Fig. 1(c). The samples to be observed are DVD and BD discs, and their protective films were removed before using. At the beginning, the diluted cell suspension was dropped onto the sample surface by using a micro tube. To prevent evaporation of the solution, a glass coverslip was used to seal the cell suspension, creating a semi-closed sample chamber. The tapered fiber was fed into the sample chamber by adjusting the manipulator. The 980 nm laser with relatively low absorption for biological specimens was used for trapping to avoid damage to the yeast cells. A power meter was used to detect the output power of the tapered fiber tip up to 5 mW before the experiment. During the experiment, we first adjusted the manipulator to position the fiber tip at desired location and then turned on the 980 nm laser for cell trapping. After firmly trapping the chain of yeast cells, the position of the imaging lenses was controlled by adjusting the manipulator to the desired imaging location.

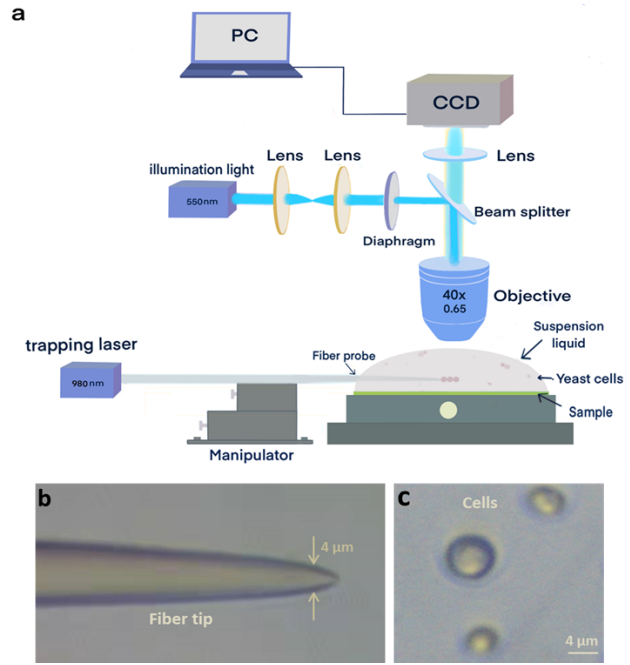

Fig. 1. Schematic experimental setup and used samples and materials. (a) Schematic diagram of the experimental setup. A conventional reflectance microscope with a CCD camera and a 40 × objective, NA=0.65, is used to observe the sample and record the images. The illumination source is a broadband LED with central wavelength of 550 nm; the fiber trapping source is a semiconductor laser with a wavelength of 980 nm. The fiber probe and the sample chamber are each attached to a micromanipulator for precision positioning. (b) A micrograph of a tapered optical fiber under optical microscope. (c) Yeast cells observed under the optical microscope.

Optical-based methods to precisely locate and arrange biological cells into ordered patterns are of great importance for biosensing and genetic engineering, etc. In addition to single cell manipulation, fiber optic tweezers can also be used to manipulate multiple biological cells simultaneously. In the experiment, yeast cells with a size of about 4 μm are selectively chosen. At about 5mW trapping power, the chosen cells can be stably trapped in a chain form. Fig. 2(a) shows the process of capturing and trapping a single yeast cell. At t=0 s, the tapered fiber was moved towards the chosen cell by adjusting the micromanipulator (Fig. 2 (a1)). Meanwhile, the trapping beam was turned on and the position of the optical fiber was adjusted to capture and trap the single yeast cell at t=3 s (Fig. 2 (a2)). Continuously adjusting the manipulator was used to ensure the cells were trapped firmly. At t=6 s, the cell was still firmly trapped at the tip of the fiber (Fig. 2 (a3)). Fig. 2(b) illustrates that different number of yeast cells can be stably trapped by the fiber tip, one after one, from zero (top) to the three cells (bottom).

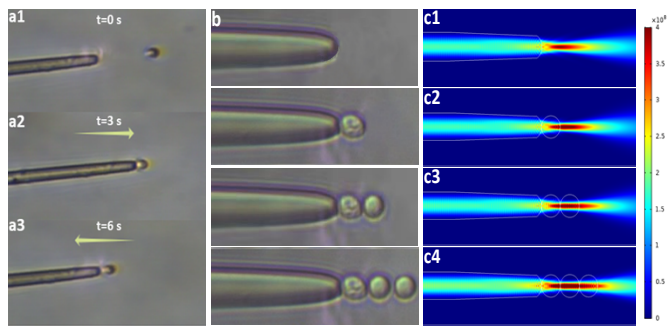

Fig. 2. Fiber optical trapping and manipulation of yeast cells. (a) Single cell (a1) At t=0s, fiber is away from the cell, (a2) at t= 3s, cell was trapped, (a3) t=6s, the cell was moved to a new location by the fiber tweezer. (b) Multiple cells (from one to three) trapped by the fiber tweezer, forming a chained cell group. (c) Numerical simulations and calculations.

The tapered shape allows the trapping beam to be highly concentrated at the fiber tip and forms a stable light trapping potential. By precisely moving the tapered fiber close to the yeast cells, the yeast cells can be captured and trapped at the tip of the fiber by the optical gradient force. Due to the spherical shape of the yeast cells, the captured laser beam can be focused to a very small area and exerts a strong optical force on other yeast cells, one after another. These cells bind together in an orderly fashion through the optical-binding effect. In order to study the capture stability of fiber optic tweezers, a theoretical model is developed using COMSOL software. Fig. 2(c) shows the energy density distribution of the 980 nm captured laser beam at the output of the bare tapered fiber. The output laser beam is focused at the tip of the fiber. When capturing a 4 μm yeast cell at the tip of the fiber, a tightly focused beam is generated (Fig. 2(c2)), which results from the interference between the field scattered by the yeast cell and the large angular component of the incident Gaussian beam through the cell. The second yeast cell is captured by using the subwavelength light spot on the front end of the yeast cell, as shown in Fig. 2(c3). There still exists the light spot on the front end of the second yeast cell, so the next cell could be trapped as well, as in Fig. 2(c4). The output light is highly focused and forms a photonic nanojet that produces a stronger optical force on the trapping yeast cell.

The imaging setup is shown in Fig. 3(a). The light-trapped yeast cells were brought close to the imaging sample surfaces (DVD and BD). The distance between sample the lens is in sub-wavelength scale so that evanescent waves can contribute to the subwavelength imaging process. Fig. 3(b) shows optical image of the grating structure on DVD disc (340 nm groove separated 400 nm apart, period: 740 nm) observed under the optical microscope without cell lens, and Fig. 3(e) for the grating structure of BD disc under the SEM (100 nm groove separated 200 nm apart, period: 300 nm) [40]. Due to optical diffraction, the surface structure of BD discs cannot be obtained under this optical microscope imaging system. The BD structures, however, can be clearly resolved by using trapped yeast cells, as shown in Fig 3(f) by two cells and Fig 3(g) by three cells. In comparison, the DVD disc, which can be resolved without cell lens, got better results when imaged by cell lenses as shown in Fig 3(c) and Fig 3(d).

The experimental results demonstrated here confirms that the chained yeast cells can indeed function as superlensing imaging lenses to resolve subwavelength feature down to 100 nm. However, this is not the point spread function (PSF) resolution for the imaging system. The calibrated PSF resolution of such system is about 240 nm following the theoretical calibration process suggested in literature [31]. However, such calibrated resolution could be misleading since the suggested calibration process relying on image contrast in obtained super-resolution image which can be easily altered by image postprocessing. The best way to claim the resolution here is to specify the 'feature resolution', but point out this is not PSF resolution as people usually refers to in conventional optical imaging system.

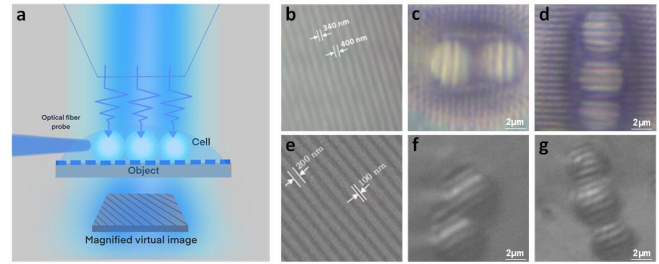

Fig. 3. Parallel subwavelength imaging by fiber-trapped cells as near-field magnifying lenses (a) Schematic of the near-field imaging setup: light trapped yeast cells were brought close to the imaging objects which were imaged and magnified by cells and projected to far-field objective. (b) DVD disc and (e) BD disc under optical microscope and SEM, respectively. (c), (f) Double-cell imaging of DVD and BD discs, respectively, (d), (g) Triple-cell imaging of DVD and BD discs.

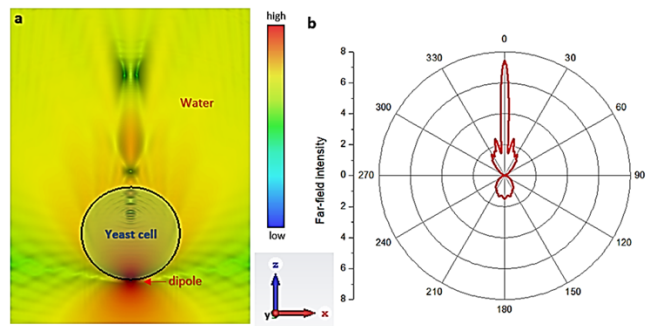

Fig. 4. Near-field imaging by yeast cell lens in water. (a) An x-polarized electric dipole source was placed 20-nm under the cell lens, the near-fields of the dipole was scattered and redirected by the yeast cell sitting on top. (b) The far-field scattering directivity showing greatly enhanced scattering in upwards direction (with cell) compared to the downwards direction (without cell).

In our work, the yeast cell was positioned extremely close to the imaging object by the trapping fiber. Near-field imaging takes place and a magnified virtual image would be generated by the cell lens and projected to the far-field objective lens. Such imaging process can be simulated by an electric dipole source located closely to the cell particle lens. Fig. 4 shows full-wave numerical simulation of such imaging process. From Fig. 4(a), we can find that the near-field waves of the x-dipole (containing evanescent sub-diffraction information) was scattered by the particle lens and directed upwards towards the microsphere objective lens. The scattering wave going downwards from the dipole source in Fig. 4(a) can be considered as the case without cell lens. From the corresponding far-field directivity map in Fig. 4(b), the energy directed upwards (with cell) is about 5 times higher than that in downward direction (without cell). This clearly demonstrate the presence of cell lenses had contributed to the conversion and collection of near-field energy with nanoscale information into the far-field, comparing to the case without cell lenses. From simulation, it can also see from Fig. 4(b) the far-field is concentrated within a collection angle of about 120-degree in upwards direction, which corresponding to numerical aperture NA=sin (60) ≈ 0.87, this would suggest the system would be able to observe smaller objects than 100 nm as in this work if we replace NA=0.65 in current study with better objective lens. This will be implemented in future work, together with the planned work on scanning imaging and image stitching to realize a non-invasive, dynamic, real-time and label-free bio-superlens scanning imaging system.

In summary, we have developed a new optical imaging device by combining fiber optical trapping technique and near-field microsphere imaging technique. By using yeast cells as imaging lenses and trapping them into a chain, we successfully demonstrated parallel imaging of 100 nm sub-diffraction features with improved efficiency and increased imaging field of view by at least three times compared to previous works. This bio-based, label-free, real-time and parallel nano-imaging device lays down a solid foundation for the development of next-generation biological super-resolution imaging devices and systems.

**Funding.** B.Y and Z.W acknowledge funding support from European Reginal Development Fund (ERDF) and Welsh Government on Center For Photonics Expertise (CPE) Project, Grant Number: 81400

**Disclosures:** The authors declare no conflicts of interest.

## References

1. L. Reimer, *Transmission Electron Microscopy. Physics of Image Formation and Microanalysis* (Transmission Electron Microscopy. Physics of Image Formation and Microanalysis, 1984).
2. E. Abbe, "Beitrage zur Theorie des Mikroskops und der mikroskopischen Wahrnehmung," Archiv Microskop. Anat. **9**, 413 (1873).
3. E. B. tzig, P. L. Finn, and J. S. Weiner, "Combined shear force and near‐field scanning optical microscopy," Applied Physics Letters **60**, 2484-2486 (1992).
4. D. Vobornik and S. Vobornik, "Scanning near-field optical microscopy," Bosn J Basic Med **59**, 63-71 (2008).
5. E. A. Ash and G. Nicholls, "Super-resolution Aperture Scanning Microscope," Nature **237**, 510-512 (1972).
6. Betzig, Eric, Patterson, George, H., Sougrat, Rachid, Lindwasser, O., and Wolf, "Imaging Intracellular Fluorescent Proteins at Nanometer Resolution," Science (2006).
7. Klar, Thomas, A., Jakobs, and Stefan, "Fluorescence microscopy with diffraction resolution barrier broken by stimulated emission," Proceedings of the National Academy of Sciences of the United States of America **97**, 8206-8206 (2000).
8. L. Schermelleh, R. Heintzmann, and H. Leonhardt, "A Guide to Super-Resolution Fluorescence Microscopy," The Journal of Cell Biology **190**, 165-175 (2010).
9. Z. Wang and B. Luk'yanchuk, "Super-resolution imaging and microscopy by dielectric particle-lenses," in *Label-Free Super-Resolution Microscopy*, V. Astratov, ed. (Springer, 2019).
10. Z. B. Wang, W. Guo, L. Li, B. Luk'yanchuk, A. Khan, Z. Liu, Z. C. Chen, and M. H. Hong, "Optical virtual imaging at 50 nm lateral resolution with a white-light nanoscope," Nat. Commun. **2**, 6 (2011).
11. L. Li, W. Guo, Y. Yan, S. Lee, and T. Wang, "Label-free super-resolution imaging of adenoviruses by submerged microsphere optical nanoscopy," Light Science & Applications **2**, 72-72 (2013).
12. H. Xiang, C. Kuang, L. Xu, H. Zhang, and Y. Li, "Microsphere based microscope with optical super-resolution capability," Applied Physics Letters **99**, 203102-203103 (2011).
13. A. Darafsheh, G. F. Walsh, L. D. Negro, and V. N. Astratov, "Optical super-resolution by high-index liquid-immersed microspheres," Applied Physics Letters **101**, 388-457 (2012).
14. A. Darafsheh, N. I. Limberopoulos, J. S. Derov, D. Jr, and V. N. Astratov, "Advantages of microsphere-assisted super-resolution imaging technique over solid immersion lens and confocal microscopies," Applied Physics Letters **104**, 061117 (2014).
15. S. Lee, L. Li, Z. Wang, G. Wei, and T. Wang, "Immersed transparent microsphere magnifying sub-diffraction-limited objects," Applied Optics **52**, 7265-7270 (2013).
16. S. Lee, L. Li, Y. Ben-Aryeh, Z. Wang, and W. Guo, "Overcoming the diffraction limit induced by microsphere optical nanoscopy," Journal of Optics **15**(2013).
17. Y. Yan, L. Li, C. Feng, W. Guo, S. Lee, and M. Hong, "Microsphere-coupled scanning laser confocal nanoscope for sub-diffraction-limited imaging at 25 nm lateral resolution in the visible spectrum," Acs Nano **8**, 1809-1816 (2014).
18. A. Darafsheh, C. Guardiola, D. Nihalani, D. Lee, J. C. Finlay, and A. Cárabe, "Biological super-resolution imaging by using novel microsphere-embedded coverslips," in *Nanoscale Imaging, Sensing, & Actuation for Biomedical Applications XII*, 2015),
19. F. Wang, S. Yang, H. Ma, S. Ping, W. Nan, W. Meng, X. Yang, D. Yun, and Y. H. Ye, "Microsphere-assisted super-resolution imaging with enlarged numerical aperture by semi-immersion," Applied Physics Letters **112**, 023101 (2018).
20. B. Yan, Z. B. Wang, A. L. Parker, Y. K. Lai, P. J. Thomas, L. Y. Yue, and J. N. Monks, "Superlensing microscope objective lens," Applied Optics **56**, 3142-3147 (2017).
21. S. Yang, Y. Cao, Q. Shi, X. Wang, T. Chen, J. Wang, and Y.-H. Ye, "Label-Free Super-Resolution Imaging of Transparent Dielectric Objects Assembled on a Silver Film by a Microsphere-Assisted Microscope," The Journal of Physical Chemistry C **123**, 28353-28358 (2019).
22. P. C. Montgomery, S. Lecler, A. Leong‐Hoï, and S. Perrin, "High Resolution Surface Metrology Using Microsphere‐Assisted Interference Microscopy," physica status solidi (a) **216**(2019).
23. F. Wang, L. Liu, H. Yu, Y. Wen, P. Yu, Z. Liu, Y. Wang, and W. J. Li, "Scanning superlens microscopy for non-invasive large field-of-view visible light nanoscale imaging," Nat Commun **7**, 13748 (2016).
24. L. A. Krivitsky, J. J. Wang, Z. Wang, and B. Luk'Yanchuk, "Locomotion of microspheres for super-resolution imaging," Scientific Reports **3**, 3-7 (2013).
25. J. Li, W. Liu, T. Li, I. Rozen, J. Zhao, B. Bahari, B. Kante, and J. Wang, "Swimming Microrobot Optical Nanoscopy," Nano Letters **16**, 6604-6609 (2016).
26. B. Yan, Z. Wang, A. Parker, Y. Lai, J. Thomas, L. Yue, and J. Monks, "Superlensing Microscope Objective Lens," Appl Opt **56**, 3142-3147 (2016).
27. J. N. Monks, B. Yan, N. Hawkins, F. Vollrath, and Z. Wang, "Spider Silk: Mother Nature's Bio-Superlens," Nano Lett **16**, 5842-5845 (2016).
28. Y. Li, H. Xin, Y. Zhang, H. Lei, T. Zhang, H. Ye, J. J. Saenz, C. W. Qiu, and B. Li, "Living Nanospear for Near-Field Optical Probing," ACS Nano **12**, 10703-10711 (2018).
29. X. Liu, J. Huang, Y. Li, Y. Zhang, and B. Li, "Optofluidic organization and transport of cell chain," J Biophotonics **10**, 1627-1635 (2017).
30. Y. Li, X. Liu, B. J. L. S. Li, and Applications, "Single-cell biomagnifier for optical nanoscopes and nanotweezers," **8**(2019).
31. K. W. Allen, N. Farahi, Y. Li, N. I. Limberopoulos, D. E. Walker, A. M. Urbas, V. Liberman, and V. N. Astratov, "Super-resolution microscopy by movable thin-films with embedded microspheres: Resolution analysis," Annalen der Physik **527**, 513-522 (2015).